\documentclass[fdp,fleqn]{D-art}
\usepackage{times}
\usepackage{D-thm}
\usepackage{psfrag}
\usepackage[]{graphicx}
\usepackage{amssymb}

\newcommand{\N}{\mathcal N}
\newcommand{\C}{\mathbb C}
\newcommand{\Z}{\mathbb Z}
\newcommand{\T}{\mathbb T}

\newcommand{\beq}{\begin{equation}}
\newcommand{\eq}{\end{equation}}

\newcommand{\drawsquare}[2]{\hbox{%
\rule{#2pt}{#1pt}\hskip-#2pt
\rule{#1pt}{#2pt}\hskip-#1pt
\rule[#1pt]{#1pt}{#2pt}}\rule[#1pt]{#2pt}{#2pt}\hskip-#2pt
\rule{#2pt}{#1pt}}

\newcommand{\fund}{~\raisebox{-.5pt}{\drawsquare{6.5}{0.4}}~}
\newcommand{\afund}{~\overline{\raisebox{-.5pt}{\drawsquare{6.5}{0.4}}}~}

\newcommand{\symm}{~\raisebox{-.5pt}{\drawsquare{6.5}{0.4}}\hskip-0.4pt%
        \raisebox{-.5pt}{\drawsquare{6.5}{0.4}}~}

\newcommand{\asymm}{~\raisebox{-3.5pt}{\drawsquare{6.5}{0.4}}\hskip-6.9pt%
        \raisebox{3pt}{\drawsquare{6.5}{0.4}}~}

\begin{document}
\DOIsuffix{theDOIsuffix}
\pagespan{1}{}



\title[Brane Tilings for Orientifolds]{Brane Tilings for Orientifolds}


\author[D. Krefl]{Daniel Krefl\inst{1,2,}%
  \footnote{E-mail:~\textsf{daniel.krefl@cern.ch}
            }}
\address[\inst{1}]{Arnold Sommerfeld Center for Theoretical Physics,\\Ludwig-Maximilians-Universit\"at, Theresienstr. 37, 80333 Munich, Germany }
\address[\inst{2}]{CERN, PH-TH Division, CH-1211 Geneva, Switzerland }

\begin{abstract}
We give an elementary introduction to the recent derivation of the effective low-energy gauge theories of D3-branes probing orientifolds of toric Calabi-Yau 3-fold singularities via brane tiling techniques.\\\\
{\it Contribution to the proceedings of the $3^{rd}$ workshop of the RTN project ÔConstituents, Fundamental Forces and Symmetries of the UniverseÕ in Valencia, 1-5 October, 2007. }

\end{abstract}
\maketitle                   





\section{Introduction}

Brane tilings, often also refered to as dimers, are currently the most powerful tool available to study the 4D low-energy effective gauge theories arising from D3-branes at toric Calabi-Yau 3-fold singularities. From a mathematical viewpoint, brane tilings are just bipartite graphs, there bipartite means that one can color the vertices of a graph with two colors (usually black and white) such that vertices of equal color are always not adjacent. Such bipartite graphs encode the effective gauge theory of D3-brane at toric Calabi-Yau singularities in a simple manner. In detail, the dictionary between such a bipartite graph and the corresponding gauge theory is as follows: 

The vertices of the graph correspond to the superpotential terms, where the signs of the terms are determined by the color of the corresponding vertices. Here we will fix the convention that black vertices correspond to negative terms. The faces of the graph represent $SU(N)$ gauge groups, while an edge separating two faces corresponds to a bifundamental field charged under both groups, i.e. transforming as $(\fund_i,\afund_j)$. The chirality of the fields are fixed by giving the edges an orientation, for example from white to black.

As an example, let us consider the simplest imaginable model, namely $N$ D3-branes probing flat-space. As is well known, the corresponding effective gauge theory is $\N=4$ SYM with the superpotential
\beq\label{introeq1}
W=\Phi_1\Phi_2\Phi_3-\Phi_1\Phi_3\Phi_2,
\eq 
where $\Phi_i$ denote the 3 adjoint fields of the single $SU(N)$ gauge factor. Note that here and in the following we always omit traces. With the above dictionary, it is easy to check that this gauge theory is encoded by the graph shown in figure \ref{fig1}a. Observe that this graph is periodic, i.e. it can be thought of as living on a $\T^2$. The fundamental cell is indicated in figure \ref{fig1}a via the centered square. This periodicity is not accidental, but rather directly linked to the conformal invariance of the gauge theory. For more details on that and further elementary aspects of brane tilings we refer to the reviews \cite{Kennaway:2007tq,Yamazaki:2008bt}. 

Instead, we shortly dwell on the reason why one can associate to gauge theories arising from D3-branes at toric Calabi-Yau 3-fold singularities a bipartite graph at all: In fact, this solely rests on the fact that these ``toric" gauge theories have the distinctive property that each bifundamental field occurs in the superpotential exactly twice, once in a term with positive sign and once in a term with negative sign. This property naturally allows to define an one-to-one map from the superpotential to a bipartite graph. A practical algorithm for that can be found in appendix A of \cite{Franco:2007ii}. One might ask why this property should hold in general for toric backgrounds. The simplest (but not fully complete) argument hinting towards that is as follows: One can see this property simply as being a left-over of the commutator structure of the superpotential occuring for $\C^3$, as is given in equation (\ref{introeq1}). In order to see that, observe that this property is conserved under orbifolding $\C^3$ to $\C^3/(\Z_M\times \Z_M)$ with $M$ an arbitrary positive integer. Since we can obtain each toric background via successive blowups from $\C^3/(\Z_M\times \Z_M)$, if we choose $M$ large enough, it remains to argue that this property is preserved as well under blowups. However, blowups correspond in the gauge theory to higgsing, i.e. giving vevs fields. Thus, the property is preserved under higgsing as well, as long as no mass terms are generated. If mass terms are generated, the property can be violated (but not necessarily). However, if the property is violated after higgsing, the performed higgsing corresponds in all known cases to a blowup to a non-toric geometry. In this sense, one should see the property really as a definition of a ``toric" gauge theory.

After this short introduction into brane tilings, let us come to the main concern of this note: Namely, the question if brane tilings encode the gauge theories arising from D3-branes probing orientifolded backgrounds as well. This would be a powerful result, since it is generally very cumbersome to obtain the effective gauge theories of D3-branes at orientifolded backgrounds via ordinary CFT techniques. 

Recall that the Type IIB orientifold action is given by $(-1)^F\omega \sigma$, where $F$ is left-moving fermion number, $\omega$ reverses the worldsheet orientation and $\sigma$ is an involution of the internal background, which we denote as $X$. In order to preserve a maximal amount of supersymmetry ($\N=1$ in most examples), $\sigma$ needs to act on the globally defined holomorphic three-form $\Omega$ of $X$ as
\beq\label{introeq2}
\sigma(\Omega)=-\Omega.
\eq
The fixed-point loci under $\sigma$ define orientifold planes, which we will denote in the following simply as $Op$-planes, where $p$ is the dimension of the corresponding fixed-point locus. The condition (\ref{introeq2}) basically just implements the fact that only O3- and O7-planes can preserve a maximum amount of common supersymmetry with the probe D3-brane.

The orientifold action on the Chan-Paton factors comes with a sign freedom such that open string states mapped to themselves are either symmetrized or anti-symmetrized. Especially, $SO(N)$ and $Sp(N)$ world-volume gauge theories and matter in the symmetric ($\symm$) and anti-symmetric (\asymm) 2-index tensor representation can occur.

This enhanced matter spectrum implies that the brane tiling concept needs to be extended to be able to capture these new ingredients. Since there is a direct connection between the brane tiling and the underlying (mirror) geometry \cite{Feng:2005gw}, it is plausible to expect that the possible geometric involutions $\sigma$ are in someway visible in the brane tiling, such that the tiling would tell us immediately the gauge group and matter identifications under the corresponding orientifold projection, which then would be just needed to be extended by some choice of Chan-Paton action. 

Indeed, recently such an extension has been obtained. In this note, we will give a short sketch of the main idea and refer the interested reader for more details and aspects to the original work \cite{Franco:2007ii}. In detail, the approach of \cite{Franco:2007ii} is to study simple models for which the orientifolded theories are known, inferring actions on the brane tiling which reproduce these known orientifolds and to extract the systematics behind, which then allows a generalization to more involved models. In the following, we will exactly follow this approach:

\section{Orientifolds acting as point reflection}

\begin{figure}
\psfrag{A}[cc][][1]{a)}
\psfrag{B}[cc][][1]{b)}
\psfrag{C}[cc][][1]{c)}
\psfrag{a}[cc][][0.5]{$\phi_1$}
\psfrag{b}[cc][][0.5]{$\phi_2$}
\psfrag{c}[cc][][0.5]{$\phi_3$}

\includegraphics[scale=0.20]{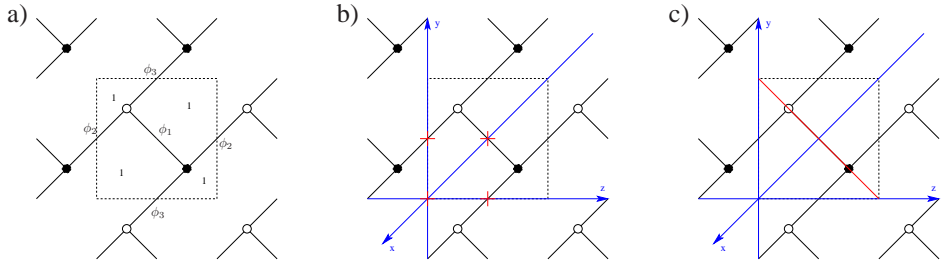}
\caption{Brane tilings encoding the gauge theories arising from $X=\C^3$ and from two different classes of orientifolds thereof. a) Parent theory, b) Orientifold acting as point reflection, c) Orientifold acting as line reflection.}
\label{fig1}
\end{figure}

Therefore, let us consider the simplest imaginable class of involutions of $\C^3$:
\beq\label{pointeq1}
\sigma(x,y,z)\rightarrow (r_1 x,r_2 y,r_3z),
\eq
with $r_i=\pm 1$. This class of involution keeps the coordinates fixed up to sign. Since in this example the globally defined holomorphic 3-form is just given by
\beq
\Omega=dx\wedge dy\wedge dz,
\eq
the supersymmetry condition (\ref{introeq2}) translates to the restriction $\prod_i r_i=-1$. Indeed, under this restriction we have only O3- or O7-planes, which can preserve a maximum amount of supersymmetry with the probe D3-branes as already stated in the introduction.

The coordinates $(x,y,z)$ of $\C^3$ correspond in the probe D3-brane world-volume gauge theory to mesonic operators which are in fact just given by the three adjoint fields $\phi_i$. Hence, from the geometric action (\ref{pointeq1}) one can easily infer the corresponding action on the brane tiling.

In order to see that, note that the mesonic operators are encoded in the brane tiling via closed loops crossing the edges corresponding to the fields the mesonic operator is composed of. The orientation of the loops are fixed by the convention that the black vertex belonging to an edge crossed by a loop is always located on the left. An example of how mesonic operators are encoded in the brane tiling can be found in figure \ref{fig1}b, where the mesonic operators are indicated as blue lines (which form closed loops in the fundamental $\T^2$). 

The involution (\ref{pointeq1}) maps the coordinates to themselves up to sign, so the involution can be seen as point reflection in the brane tiling. Especially, the fundamental $\T^2$ has 4 fixed-points under the reflection.\footnote{Note that consistency requires that the point-reflection always identifies vertices of different color \cite{Franco:2007ii}.} This is indicated in figure \ref{fig1}b via the red crosses. Observe that in our $\C^3$ example the point-reflection in the brane tiling maps the single face and the three adjoint fields to itself, while the two superpotential terms are identified. This is as expected from field theory considerations. However, we still need to find a way to determine the possible Chan-Paton actions on the spectrum, i.e. which combinations of enhanced gauge groups and (anti)-symmetric representations are allowed. For that, we assign signs $c_i=\pm 1$ to the 4 fixed-points, with the rule that if a fixed point with positive sign sits on a face, the gauge group corresponding to the face is projected to $SO$, while for a negative sign to $Sp$. If a fixed point with positive sign sits on an edge, the corresponding field is projected to the symmetric representation, while it is projected to the anti-symmetric representation if the fixed point has negative sign. However, things are more complicated since we apparently have a mixture of the geometric action and the Chan-Paton action, i.e. the 4 fixed-points must encode the allowed signs $r_i$ of the action (\ref{pointeq1}) as well (especially, this means that the $c_i$ do not necessarily correspond to physical charges). That is, the mesonic operators should pick up a sign from the fixed-points they are passing through (for which explicit rules can be defined, for details see \cite{Franco:2007ii}). However, since the $r_i$ need to fulfill the constraint $\prod_i r_i=-1$, this immediately tells us that there must be also a constraint on the $c_i$ to implement the supersymmetry condition (\ref{introeq2}). 

We postulate that this global constraint on the $c_i$ is generally given by 
\beq\label{pointeq2}
P :=\prod_i c_i=1-2 \left(\frac{N}{2} \mod 2\right),
\eq
where $N$ is the number of superpotential terms. For our $\C^3$ example this just tells us that $P$ needs to be odd. Using odd $P$, we can indeed reproduce all supersymmetry preserving orientifolds of $\C^3$. For example, $c_i=(-+++)$ gives a $Sp$ group with 3 symmetrics, denoted as $S_i$, and superpotential $W=S_1S_2S_3$ (which is generally obtained by replacing the fields in the identified superpotential terms by their orientifold images).

Since we can obtain from $\C^3$ via orbifolding and higgsing essentially all toric backgrounds, as sketched in the introduction, it is expectable that the above systematics, i.e. that orientifolds keeping the coordinates fixed up to sign act as point reflection on the brane tiling with Chan-Paton sign assignment restricted by the constraint (\ref{pointeq2}), are valid in general. However, for performing the higgsing, one should keep in mind that only blowups are allowed that are compatible with the involution $\sigma$. In brane tiling language, this restricts the edges allowed to be removed to edges fixed under the point-reflection or to a pairwise removal of edges (which are mirror to each other).

\begin{figure}
\psfrag{a}[cc][][1]{a)}
\psfrag{b}[cc][][1]{b)}
\psfrag{c}[cc][][1]{c)}
\includegraphics[scale=0.20]{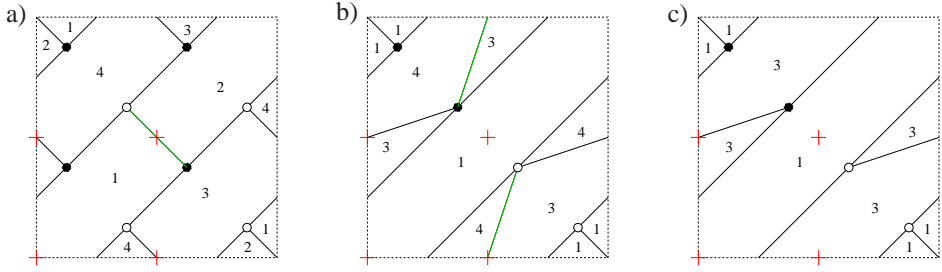}
\caption{Brane tilings for orientifolds of  a) $\C^3/(\Z_2\times \Z_2)$, b) $SPP$, c) $\C^2/\Z_2\times \C$. The green edges are the ones to be removed in order to perform the higgsing (blowup) $\C^3/(\Z_2\times \Z_2)\rightarrow SPP\rightarrow \C^2/\Z_2\times \C$ performed in \cite{Park:1999ep}.}
\label{fig2}
\end{figure}

Indeed, using the above systematics we can reproduce essentially all known orientifold models with $\sigma$ keeping the coordinates fixed up to sign in a simple fashion. For example, from the brane tilings shown in figure \ref{fig2} one can directly infer the orientifolded theories originally derived via field theory techniques in \cite{Park:1999ep}. 

Finally, note that anomaly cancellation often restricts the allowed ranks of the gauge groups and/or requires the introduction of extra fundamental fields due to additional non-compact D7-branes, which have to be added for a consistent model. Recently, this has been discussed in some more detail in \cite{Imamura:2008fd}.

\section{Orientifolds acting as line reflection}
Let us now consider another class of possible involutions acting on $\C^3$:
\beq
\sigma(x,y,z)\rightarrow (r_1\tau(x),r_2\tau(y),r_3\tau(z)),
\eq
where $\tau$ is a permutation which exchanges two of the coordinates. Clearly, this class satisfies (\ref{introeq2}), if $\prod_i r_i=+1$. As expected, under this restriction the fixed-point loci correspond to O3- or O7-planes.
As before, we translate this geometric action to an action on the brane tiling via the mesonic operators corresponding to the coordinates. This is illustrated in figure \ref{fig1}c. We infer that this class of orientifolds acts as a line reflection on the brane tiling.\footnote{Note that consistency requires that the line reflection always identifies vertices of equal color \cite{Franco:2007ii}.} In comparison with the point reflection, we have instead of fixed points now one or two fixed lines, depending on the geometry of the fundamental cell. Similar as for the fixed points, the fixed lines will tell us the projected spectrum, i.e. as before we assign signs $c_i$ to the fixed lines which stand for the two possible Chan-Paton actions. However, since two of the coordinates are mapped to each other, the corresponding two mesonic operators in the brane tiling will have exactly the same intersection pattern with the fixed lines, and thus we can reproduce with the brane tiling only orientifolds with fixed relative sign of the $r_i$ of the exchanged coordinates. This in turn tells us that the remaining coordinate should not pick up a sign in order to preserve a maximal amount of supersymmetry. However, since the mesonic operator that is mapped to itself crosses the fixed-line exactly two times in our $\C^3$ example, it seems to be plausible that this is automatically satisfied, hence we do not have a global constraint of the form (\ref{pointeq2}), as in the point-reflection case.

Since we can as before obtain all other toric models by orbifolding and higgsing $\C^3$ in a proper way, it is expectable that the global constraint is generally absent for this class of orientifolds. Indeed, the further examples of line reflections considered in \cite{Franco:2007ii} do not require such a constraint. 

\begin{figure}
\psfrag{a}[cc][][1]{a)}
\psfrag{b}[cc][][1]{b)}
\psfrag{c}[cc][][1]{c)}
\psfrag{S}[cc][][0.7]{Seiberg duality}
\includegraphics[scale=0.20]{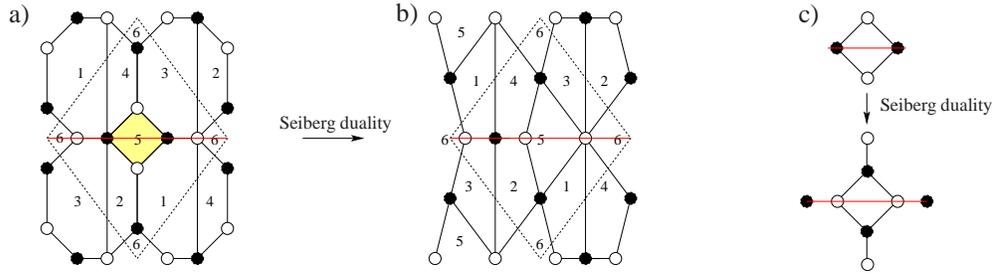}
\caption{Orientifolds acting as line 
reflection on the brane tilings of a) $dP_3^I$ and b) $dP_3^{II}$. The orientifolds of both phases are related via Seiberg duality, for instance via dualizing face 5 as sketched in c).}
\label{fig3}
\end{figure}

As further example, let us consider orientifolds of (the cone over) $dP_3$ acting as line reflection. For that, recall that $dP_3$ comes in four toric phases which are Seiberg dual to each other. However, only two of the phases, namely $dP_3^I$ and $dP_3^{II}$ apparently have a  proper $\Z_2$ symmetry, i.e. one that identifies only equal color vertices. Brane tilings for these two phases and orientifolds thereof are shown in figure \ref{fig3}a and \ref{fig3}b. Observe that the rule of how to perform Seiberg duality (for $SU(N)$ groups) in brane tilings \cite{Franco:2005rj} can be extended to the orientifoded case as well. In detail, recall that one performs Seiberg duality to an other toric phase in brane tilings by acting on squared faces as shown in figure \ref{fig3}c. Clearly, for square faces invariant under the $\Z_2$ reflection, this is automatically compatible with the orientifold projection. In fact, this just implements the well known dualities for $SO(N)$ and $Sp(N)$ groups with matter in the fundamental representation. For a square face away from the fixed-line, one can perform Seiberg duality as in the non-orientifolded case, however one needs to keep in mind that in the "covering space" brane tiling one needs to dualize simultaneously the mirror face as well. 

More interesting is the case if a face which is to be dualized possesses an edge which is fixed under the orientifold action. In this case, it is necessary to consider Seiberg duality for theories with 2-tensor representations, which is more involved and it is not clear if there are (at least for some setups) toric duals. It would be interesting to clarify this further. 

Finally, note that one has still to impose anomaly cancelation and hence in some models one might need to introduce extra flavor branes and/or choose specific ranks. Further, there is still the question of conformality, i.e. it is not apparent if it is always possible to arrange for a conformal setup, or if there are cases there a non-trivial RG-flow can not be avoided. We leave this as an open issue for further investigations.

\section{A particular application}
\begin{figure}
\psfrag{e}[cc][][1]{a)}
\psfrag{f}[cc][][1]{b)}
\psfrag{a}[cc][][0.5]{$\lambda$}
\psfrag{b}[cc][][0.5]{$\lambda'$}
\psfrag{c}[cc][][0.5]{$\phi$}
\psfrag{S}[cc][][0.7]{Seiberg duality}
\psfrag{g}[cc][][0.5]{$q$}
\psfrag{h}[cc][][0.5]{$q'$}

\includegraphics[scale=0.20]{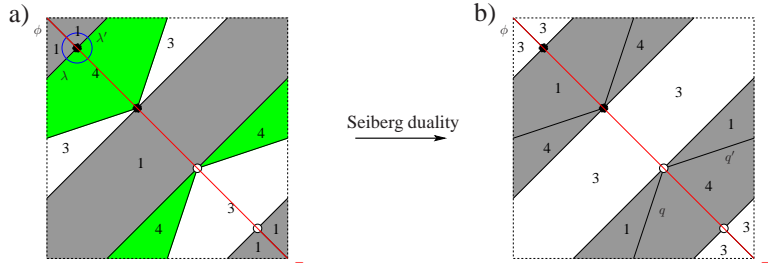}
\caption{a) A line orientifold of $SPP$ with face 4 (green) corresponding to a ``stringy" D-instanton, face 1 (grey) to an ordinary gauge group and face 3 (white) empty. b) Performing Seiberg duality on face 4 transforms the ``stringy" instanton to a gauge theory strong coupling effect.}
\label{fig4}
\end{figure}

Besides giving an easy and powerful way to obtain the effective low-energy gauge theory of D3-branes at orientifolded toric singularities, there is another interesting application of brane tilings.
Namely, from brane tilings one can as well easily infer the possible contributions of  ``stringy" D-instantons. In order to see that, recall that a ``stringy" D-instanton corresponds to an euclidean D-brane wrapped on a cycle which is not occupied by another D-brane which fills space-time as well. It is clear, that in brane tiling language there is no big difference between D-instantons and ordinary D-branes, i.e. the faces encode both. Thus, one can use the brane tiling to directly infer the charged fermionic zero-modes, i.e. strings stretching between the D-instanton brane and an ordinary D-brane and the couplings between them. In detail, the possible couplings are determined by the superpotential terms (corresponding to the vertices). This is obvious from the mirror IIA geometry, since the superpotential terms and as well the possible instanton fermionic zero-mode couplings have the same origin, namely disk instantons: If a cycle is occupied by a space-time filling D-brane, the disk instanton ending on the brane will yield a space-time superpotential. If however the cycle is occupied by a ``stringy" instanton, the same disk instanton will yield a fermionic zero-mode coupling. Recently, this fact has been also discussed in \cite{Argurio:2008jm,Kachru:2008wt}. An example of a ``stringy" D-instanton in an orientifold of the $SPP$ geometry is shown in figure \ref{fig4}. 

The orientifold which acts as line reflection on the brane tiling projects the world-volume gauge theory of the ``stringy" D-instanton corresponding to face 4 to $O(1)$, such that a contribution is possible, i.e. we have only two fermionic zero-modes in the uncharged sector. The blue circle indicates the coupling which absorbes the charged fermionic zero-modes $\lambda$ and $\lambda'$: $\lambda\phi\lambda'$ in covering space and $\lambda A\lambda$ in quotient space, where $A$ is an anti-symmetric 2-tensor arising from the projection of the adjoint $\phi$. Hence, the ``stringy" instanton will lead to a non-perturbative superpotential of the form 
\beq\label{insteq1}
W_{np}\sim Pf(A). 
\eq
Let us perform Seiberg duality on face 4. This can be done as illustrated in figure \ref{fig3}c and will yield the brane tiling shown in figure \ref{fig4}b (we take the rank of the gauge group corresponding to face 1 to be $2N$). Then, face 3 corresponds to a $Sp(2(N-2))$ gauge group which has $2N$ quarks (hence $N$ flavors) given by the bifundamental $q$. It is well known that a $Sp(2N_c)$ group with $N_f=N_c+2$ undergoes s-confinement and possesses a non-perturbative superpotential of the form (\ref{insteq1}). Thus, we have dualized the ``stringy" instanton to strong coupling dynamics, as first observed in \cite{Aharony:2007pr}. This is not accidental, but comes from the fact that this kind of ``stringy" instanton is the UV completion of an ordinary gauge theory instanton of a completely broken gauge group, as discussed in \cite{Krefl:2008gs}.

\begin{acknowledgement}
It is a pleasure to thank S. Franco, A. Hanany, J. Park, A. M. Uranga and D. Vegh for collaboration on the presented results and I. Garcia-Etxebarria for related discussions. Further, I like to thank the organizers of the $3^{rd}$ RTN workshop for the opportunity to present this work. 
This work is supported by an EU Marie-Curie EST fellowship.
\end{acknowledgement}

\end{document}